\documentclass[twocolumn,pre]{revtex4-1}
\usepackage{color}
\usepackage{graphics,graphicx,epsfig,wrapfig}
\usepackage{epsf,epstopdf}
\usepackage{amssymb,amsfonts,amsmath}
\usepackage{enumitem}
\usepackage{setspace}

\usepackage{ifthen}

\newcommand\mydots{\hbox to 0.8em{.\hss.\hss.}}
\newcommand{\beqn}{\begin{eqnarray}}
\newcommand{\eeqn}{\end{eqnarray}}
\newcommand{\beq}{\begin{equation}}
\newcommand{\eeq}{\end{equation}}

\newcommand{\ff}{1}

\newcommand{\mytitle}{Detecting T-cell receptors involved in immune responses from single repertoire snapshots }
\newcommand{\myauthors}{Mikhail V. Pogorelyy$^{1,2}$, Anastasia A. Minervina$^{1}$, \\Mikhail Shugay$^{1-5}$,
Dmitriy M. Chudakov$^{1-5}$, Yuri B. Lebedev$^{1,6}$,\\
Thierry Mora*$^{7}$, Aleksandra M. Walczak*$^{8}$}

\makeatletter
\def\@seccntformat#1{%
  \expandafter\ifx\csname c@#1\endcsname\c@section\else
  \csname the#1\endcsname\quad
  \fi}
\makeatother

\begin{document}
\title{\mytitle}
\author{\myauthors}
\affiliation{~\\
\normalsize{$^{1}$ Shemyakin-Ovchinnikov Institute of Bioorganic
  Chemistry,}
\normalsize{Moscow, Russia}\\
\normalsize{$^{2}$Pirogov Russian National Research Medical University, Moscow, Russia}\\
\normalsize{$^{3}$Privolzhsky Research Medical University, Nizhny Novgorod, Russia}\\
\normalsize{$^{4}$Center of Life Sciences, Skoltech, Moscow, Russia}\\
\normalsize{$^{5}$Masaryk University, Central European Institute of Technology, Brno, Czech Republic}\\
 \normalsize{$^{6}$Moscow State University, Moscow, Russia}\\
\normalsize{$^{7}$ Laboratoire de physique statistique,}
\normalsize{CNRS, Sorbonne Universit\'e, Universit\'e Paris-Diderot, and \'Ecole normale sup\'erieure (PSL University), Paris,
  France}\\
\normalsize{$^{8}$ Laboratoire de physique th\'eorique,}
\normalsize{CNRS, Sorbonne Universit\'e, and \'Ecole normale sup\'erieure (PSL University), Paris, France}\\
\normalsize{\rm *These authors contributed equally.}\\
}

\begin{abstract} 
{
Hypervariable T-cell receptors (TCR) play a key role in adaptive immunity, recognising a vast diversity of pathogen-derived antigens. High throughput sequencing of TCR repertoires (RepSeq) produces huge datasets of T-cell receptor sequences from blood and tissue samples { \cite{Heather2017,Lindau2017}}. However, our ability to extract clinically relevant information from RepSeq data is limited, mainly because little is known about TCR-disease associations.
Here we present a statistical approach called ALICE (Antigen-specific Lymphocyte Identification by Clustering of Expanded sequences) that identifies TCR sequences that are actively involved in the current immune response from a single RepSeq sample, and apply it to repertoires of patients with a variety of disorders --- autoimmune disease (ankylosing spondylitis \cite{Komech2018}), patients under cancer immunotherapy \cite{Robert2014,Subudhi2016}, or subject to an acute infection (live yellow fever vaccine \cite{Pogorelyy2018_yf}).
The method's robustness is demonstrated by the agreement of its predictions with independent assays, and is supported by its ability to selectively detect responding TCR in the memory but not in the na{\"\i}ve subset.
ALICE requires no longitudinal data collection \cite{Pogorelyy2018_yf,DeWitt2015} nor large cohorts \cite{Emerson2017,Faham,Pogorelyy2018_pgen}, and is thus directly applicable to most RepSeq datasets. Its results facilitate the identification of TCR variants associated with a wide variety of diseases and conditions, which can be used for diagnostics, rational vaccine design and evaluation of the  adaptive immune system state.
}
\end{abstract}

\maketitle

 \begin{figure*}
\noindent\includegraphics[width=\ff\linewidth]{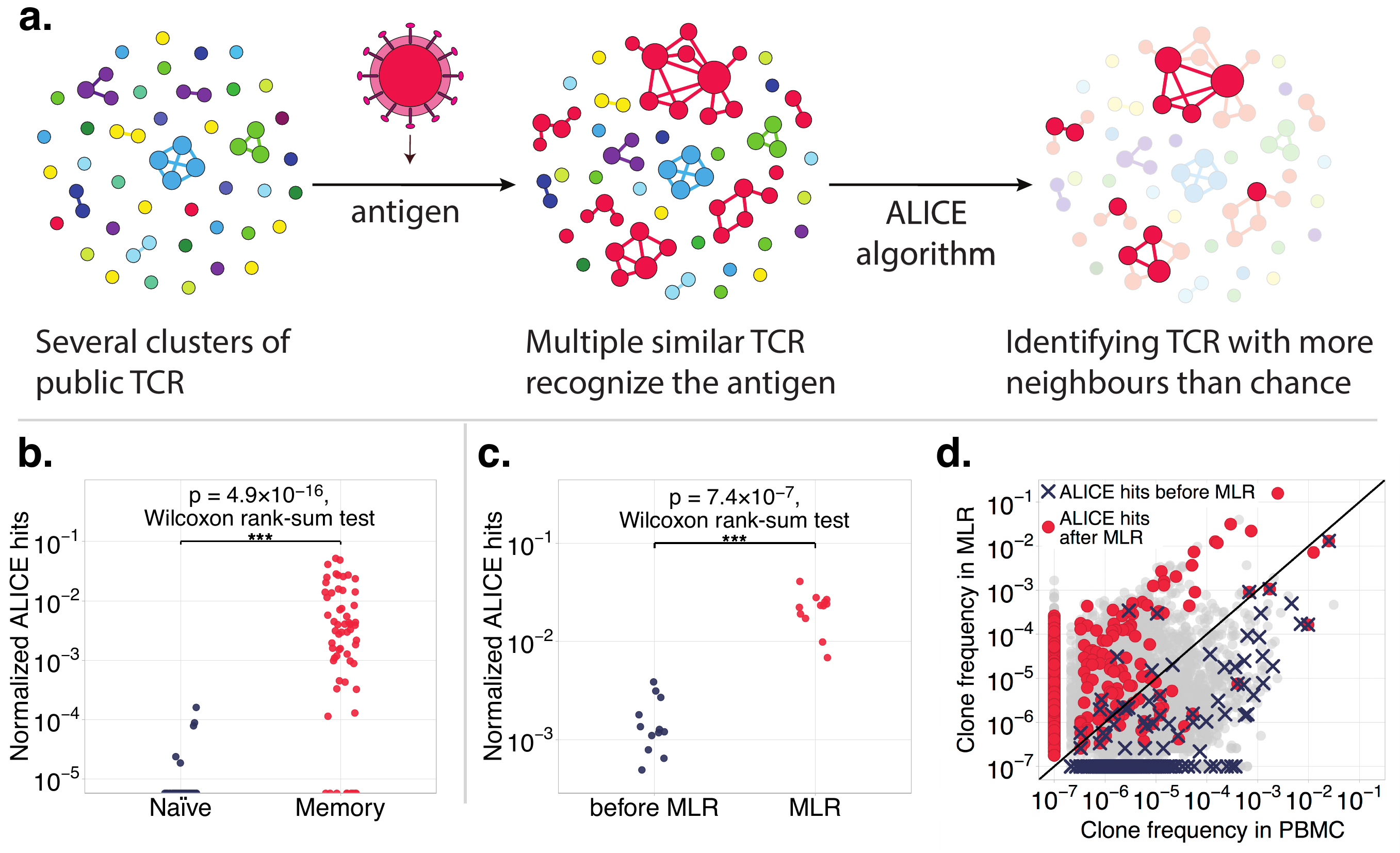}
\caption{{\bf Identification of antigen-responding clonotypes using their recombination-based frequency. a,} ALICE identifies locally enriched regions of the TCR sequence space, represented here as a graph. Vertices are TCR clonotypes observed in the repertoire, and edges connect sequences differing by at most one CDR3 amino acid. Antigen exposure induces the proliferation of multiple clonotypes with similar sequences recognizing a few immunodominant epitopes. ALICE identifies clonotypes with a higher numbers of neighbours than expected by a null model of recombination, separating clusters of antigen-responding clonotypes (in red) from clusters arising from recombination statistics (blue, green and purple clusters). {\bf b,} The number of significant results (normalized by the total number of unique nucleotide sequences) found in na{\"\i}ve vs memory published TCR beta repertoires \cite{Thome2016} demonstrates ALICE's ability to selectively detect immune response signatures in the memory subset only. {\bf c,} Normalized number of significant hits found in published repertoires of MLR cultures compared to an unstimulated control \cite{Emerson2014}. The algorithm finds many more hits in the MLR repertoire. {\bf d,} Most clonotypes identified as antigen-responding in MLR culture expanded during the assay, as evidenced by their higher frequency in MLR culture than in the control (red dots). By contrast, clonotypes identified in the unstimulated repertoire (blue crosses) mostly remain unexpanded after the assay, as they probably are signatures of previous immune responses.
}
\label{fig1}
\end{figure*}

Recent work has shown that TCR recognizing the same epitopes often have similar sequences \cite{Glanville2017,Dash2017,Shugay2017,Venturi2006,Qi2016}. However, highly similar TCR may also arise regardless of their binding properties, by virtue of their high generation probability by V(D)J recombination \cite{Venturi2011,Elhanati2018}. Indeed, clusters of similar TCR are found even in na{\"\i}ve repertoires \cite{Glanville2017,Madi2017}. Rather than relying on the number of similar sequences {\em per se}, we propose to evaluate that quantity relative to the baseline expectation from V(D)J recombination statistics to identify clusters of TCR responding to the same antigen (as schematized in Fig.\,1a).

For each TCR amino acid sequence in the data ALICE uses a stochastic TCR recombination model \cite{Murugan2012,Marcou2017} to estimate the fraction of the repertoire composed of TCR variants, called `neighbours', differing by at most one amino acid in their Complementarity Determining Region 3 (CDR3) \cite{Pogorelyy2018_pgen}. This allows us to predict theoretically the number of neighbouring clonotypes (nucleotide sequences) for each TCR under the null hypothesis of no antigen-driven TCR selection, and identify TCR with a significantly higher number of neighbours in the data than the null expectation (Online Methods). We refer to such significant results as ALICE signatures or hits. Although the basic version of the algorithm discards clonotype abundances and should thus be sensitive to sequencing depth, we also implemented an advanced (but much slower) version that includes read counts and shuffles them among clonotypes in the null (Online Methods).

As a minimal requirement for its validity, we applied our algorithm to published na{\"\i}ve (CD45RA+CCR7+) and effector memory (CD45RA-CCR7-) TCR beta repertoires from \cite{Thome2016}. Our algorithm identified multiple signatures in the memory subsets, and virtually no significant hits in the na{\"\i}ve subsets (Fig.\,1b and Fig.\,S1), in agreement with the definition that na{\"\i}ve cells have never responded to antigen stimulation.

To further validate the method's ability to detect clonal expansion during an ongoing immune response, we applied it to published TCR beta repertoires from mixed lymphocyte reaction (MLR) assay \cite{Emerson2014}. In this assay, peripheral blood mononuclear cells (PBMC) from two individuals (a responder and a stimulator) are mixed, and reactive T-cell clones from the responder's repertoire proliferate in response to the antigens presented by the stimulator's cells. ALICE identified many more hits in the responder's repertoire in the MLR culture than in unstimulated cells (Fig.\,1c). Further, the clonotypes identified by ALICE are enriched in MLR culture compared to bulk PBMC (Fig.\,1d), clearly demonstrating that these hits correspond to antigen-specific clonal expansions.

\begin{figure*}
\noindent\includegraphics[width=\linewidth]{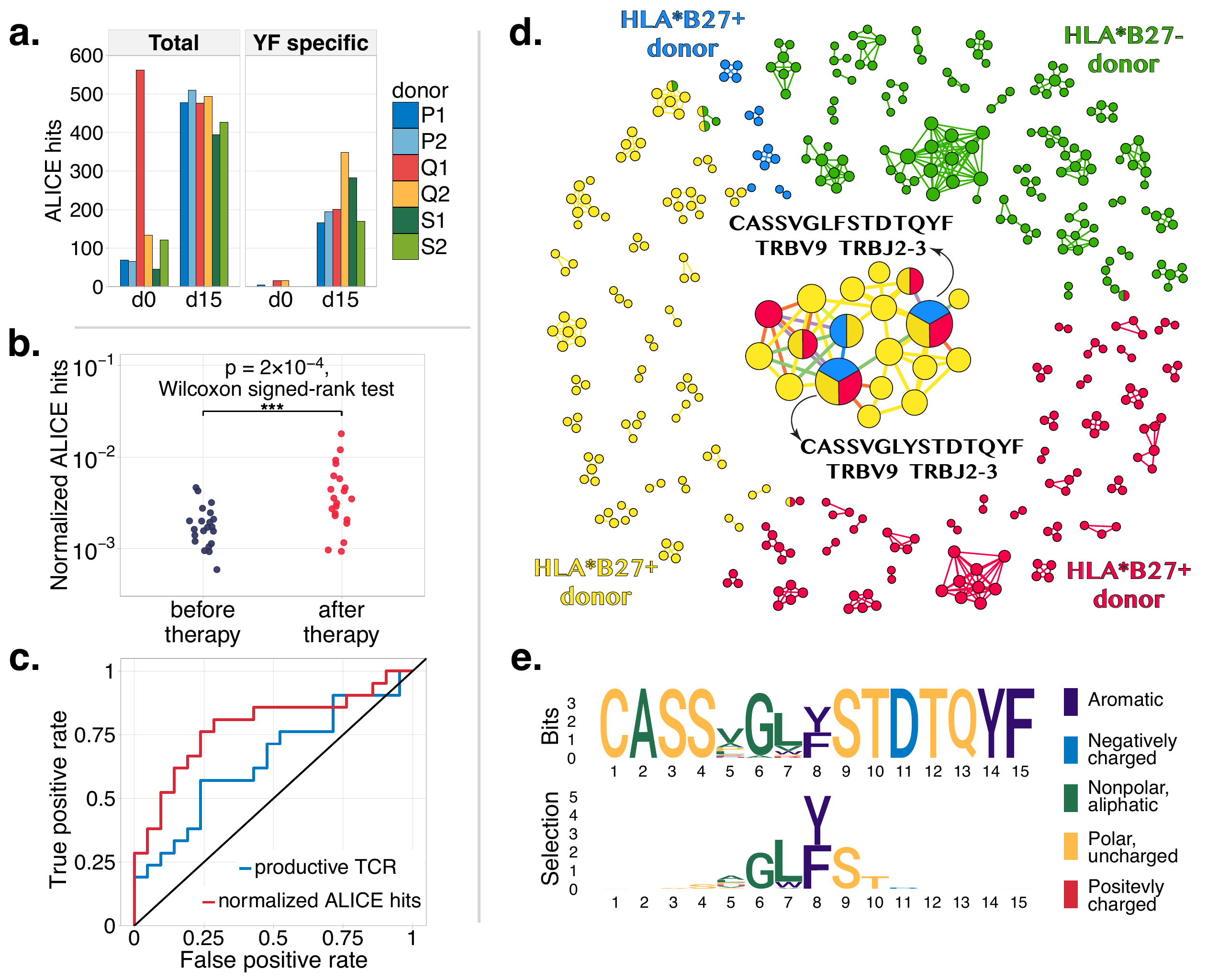}
\caption{{\bf a, Identification of reactive clonotypes following immunization.} Left panel shows the total number of ALICE immune response signatures before (day 0) and on the peak of the response to yellow fever (YF) vaccine (day 15). Right panel shows the number of clonotypes identified by the algorithm that have high similarity to known YF-reactive clonotypes from \cite{Pogorelyy2018_yf}. {\bf b, Analysis of peripheral blood repertoires before and after checkpoint blockade cancer immunotherapy \cite{Robert2014}.} The larger number of ALICE signatures after (red) than before (blue) therapy represent clones triggered by the therapy. {\bf c,} Receiver operating characteristic (ROC) curves for distinguishing pre- and post immunotherapy repertoires. The number of ALICE hits (red, area under the ROC, AUROC=0.78) is a better discriminant than the number of unique clonotypes (blue, AUROC=0.65). {\bf d, Graph of expanded clonotypes in synovial fluid of 4 ankylosing spondylitis (AS) patients.} Vertices represent significant clonotypes identified by the algorithm, and edges connect clonotypes with at most one amino acid mismatch. Zero-degree vertices are not shown. Vertices are colored according to the patients, and split vertices represent public sequences identified in several donors. The two sequences shared among all 3 HLA-B27+ patients were previously associated with AS and HLA-B27. {\bf e,} While the classical sequence logo of the central cluster in {\bf d} is dominated by germline-encoded positions (top), selection factors highlight position specific pressures acting on the expanded sequences (bottom). 
}
\label{fig2}
\end{figure*}

We then asked whether our method could identify TCR specific to a particular target using an {\em in vivo} acute viral infection model. In a previous study, peripheral blood of 6 donors was collected and their TCR beta repertoire sequenced at several time points before and after live yellow fever vaccine (YF-17D) immunization \cite{Pogorelyy2018_yf}. Clonotypes that significantly expanded following vaccination were identified by temporal comparisons. Here we applied ALICE to each time point to identify expanded clonotypes independently, using only single repertoire snapshots.

ALICE identified more immune response signatures on the peak of response (day 15) than before immunization (day 0) in all donors (Fig.\,2a, left), except one who was probably undergoing another immune response at the moment of immunization (see Fig.\,S2). Applying the advanced version of ALICE with read counts yielded almost identical results (Fig.\,S3). To validate ALICE's expanded clonotypes, we compared its predictions to known YFV17D reactive sequences obtained from longitudinal data in the source study. We found that 40-70\% of ALICE hits on day 15 were highly similar (same VJ usage and up to one CDR3 amino acid mismatch) to previously identified yellow fever specific clonotypes (Fig.\,2a, right).

Next we applied our approach to peripheral blood TCR beta repertoire samples from two CTLA4 checkpoint blockade cancer immunotherapy studies \cite{Robert2014,Subudhi2016}. We found more ALICE immune response signatures after the treatment than before (Fig.\,2b and Fig.\,S4). The number of these signatures is a better measure than previously proposed summary statistics of peripheral TCR repertoires aimed at detecting the effect of immunotherapy:
it discriminated pre- and post-treatment time points better than the number of unique clonotypes (richness) proposed in \cite{Robert2014}, as evidenced by the comparison of their receiver operating characteristic curves (Fig.\,2c).
ALICE hits increased more significantly after immunotherapy than richness or Shannon entropy (a classical measure of diversity), both in the data from \cite{Robert2014} ($p=0.0002$ for ALICE vs. $p=0.0005$ for richness and $p=0.04$ for entropy, Wilcoxon signed rank test), and in the data from \cite{Subudhi2016} ($p=0.0003$ for ALICE vs. $p=0.6$ for richness and $p=0.6$ for entropy).
By contrast to these other measures, ALICE identifies particular clonotypes which are likely to be activated by the therapy. Tracking of such clonotypes in time and in the tumor tissue could provide insights into therapy efficiency and adverse effects.

Lastly, we asked if ALICE was able to identify condition-associated clonotypes in patients with autoimmune diseases. We analysed four TCR beta repertoires of CD8+ T cells from the synovial fluid of ankylosing spondylitis (AS) patients from \cite{Komech2018}. Fig.\,2d shows clusters of ALICE-predicted clonotypes in 3 HLA-B27+ donors and one HLA-B27- donor. Although most predicted TCR were patient specific, 2 clonotypes were independently found in all 3 HLA-B27+ patients, but not in the HLA-B27- patient. These two clonotypes exactly coincide with previously reported public clonotypes in a population of HLA-B27+ patients with AS \cite{Komech2018,Faham}, and also found in synovial fluid spectratyping of patients with AS and reactive arthritis \cite{Dulphy1999,May2002}. The independent identification of these sequences by ALICE demonstrates the relevance of its predictions, and suggests that these public clonotypes actively participate in the immune response in inflamed joints. ALICE also predicts previously unreported patient specific expanded clonotypes, which population studies cannot detect by design.

To visualize CDR3 sequence motifs identified by the algorithm, we developed a novel approach to highlight differences in amino acid composition relative to the background recombination statistics (similar to \cite{Dash2017}), based on a position-weight matrix selection model learned from the TCR sequence subset (\cite{Elhanati2014}, see Online Methods). In a classical sequence logo derived from the central cluster of Fig.\,2d, positions encoded in the germline by V and J segments at the two ends of the CDR3 are very conserved and dominate the logo (Fig.\,2e, top). By contrast, our selection logo highlights amino acids that are enriched relative to that baseline (Fig.\,2e, bottom), showing a high enrichment in aromatic (Y and F) residues at CDR3 position 8. We speculate that these residues form contacts with the antigen that are crucial for TCR recognition.

By robustly identifying expanded clonotypes {\em in vivo} from a single RepSeq dataset without any information about the antigens of interest, ALICE overcomes a major limitation of MHC-multimer staining \cite{Davis2011,Glanville2017,Dash2017,Shugay2017} --- a popular {\em ex vivo} assay commonly used for identifying condition-associated TCR sequences --- which requires prior knowledge of the antigenic peptide and HLA allele {presenting it}. Our method can thus be applied to variety of conditions for which this information in not available, from autoimmune disease to infection models. TCR sequences can also be associated with certain conditions without antigenic information by sequencing huge cohorts of patients \cite{Emerson2017,Pogorelyy2018_pgen,Faham,Komech2018,DeWitt2018}, but these methods are very costly. ALICE could leverage the statistical power of such large cohort studies if combined with the computational method of \cite{Pogorelyy2018_pgen}.

Our approach have several limitations. It can only identify responding TCR with high enough frequencies. A significant fraction of responding TCR are individual specific {\cite{Pogorelyy2018_yf,Madi2017,Qi2016}}, and are unlikely to have similar variants and thus to be detected by the algorithm. Extending our method to more refined distance measures (e.g. \cite{Dash2017}) could help mitigate this issue. Another limitation is a natural consequence of its main advantage -- antigen independence. In individuals with multiple conditions, the algorithm will identify clonotypes associated with all of them, with no way of telling them apart. Repeating the analysis at different time points (e.g. after clearance of an infection) can help to distinguish TCR associated with each condition. In Fig.\,S1 we performed such an analysis for the outlying yellow fever vaccinee Q1, who was probably undergoing another transient immune response, and identified ALICE signatures that were truly yellow fever specific.

The particular sequences identified by ALICE can be studied across patients that share a condition to identify publicly responding clones,  as we did in the ankylosing spondylitis example. They could be tracked over time during and after the disease to help design biomarkers for diagnostics and to understand the persistence of immune memory. They could be searched in the repertoires of T-cell subpopulations to gain insight into their immunological function. As more repertoire sequence datasets associated with various conditions are being collected, ALICE could be used to rapidly grow databases of condition-specific TCR, with applications in the diagnostic and treatment of diseases.

\section*{Methods}

{\bf ALICE statistical model formulation.} The algortihm operates on a dataset of $n$ unique nucleotide TCR sequences (clonotypes) with a given VJ combination. The procedure is then applied to all VJ combinations present in the data. Unique nucleotide sequences have corresponding amino acid sequences. The goal is to find outlying sequences that have an abnormal number of nucleotide variants in the data that differ by at most one amino acid.

For each amino acid sequence $\sigma$, under the null hypothesis we expect the number of neighbours $d$ to be Poisson distributed:
\begin{equation}
P(d|\sigma)=e^{-\lambda}\frac{\lambda^d}{d!}, \label{eq:pois}
\end{equation}
with mean $\lambda=n \sum_{\sigma'\sim \sigma}Q P_{\rm gen}(\sigma')$. The sum is over all possible similar variants $\sigma'$ of $\sigma$. Here, similarity $\sigma'\sim\sigma$ is defined by having at most one amino acid mismatch, but other measures could be used instead.
$P_{\rm gen}(\sigma')$ is the probability to generate a given amino acid sequence $\sigma'$ by V(D)J recombination, and $Q$ a rescaling factor accounting for thymic selection \cite{Pogorelyy2018_pgen,Elhanati2018} which eliminiates a fraction $1/Q$ of generated sequences. Its value was set to $Q=9.41$ as the average over all VJ combinations reported in \cite{Pogorelyy2018_pgen}.

{\bf Estimating the generation probability of amino acid sequences.} We estimated $P_{\rm gen}(\sigma)$ of amino acid sequences by Monte-Carlo simulation, as described in \cite{Pogorelyy2018_pgen}. We generated 200 millions TCR with fixed VJ choice {\em in silico} using the VDJ recombination model from \cite{Murugan2012}. Sequences were then translated to amino acids, and the overall frequency of each distinct amino acid sequence was estimated by counting. 
The advantage of Monte Carlo simulations is that it can be done for all sequences of interest simultaneously. For datasets of moderate size, exact computation of $P_{\rm gen}$ of each sequence of interest, e.g. by OLGA \cite{Sethna2018}, may be faster.
Another option would be to use large number of published datasets \cite{Emerson2017,Britanova2016}, and treat the number of occurrences of each TCR sequence of interest in these datasets as a proxy for TCR recombination probability, as implemented into VDJtools \cite{Shugay2015}. An implementation of the corresponding routine for VDJtools software framework is described at \url{http://vdjtools-doc.readthedocs.io/en/master/annotate.html#calcdegreestats}. Note that VDJtools implementation allows setting an arbitrary Levenstein distance threshold for defining neighboring clonotypes. Forcing clonotypes to have the same VJ/V segments or allowing segment mismatches is also optional. The implementation relies on a pre-compiled control dataset instead of using a generative VDJ rearrangement model, control datasets can be obtained from \url{https://zenodo.org/record/1318986}.

{\bf ALICE pipeline. } Nucleotide sequences with low numbers of reads (here, singletons) were first removed to get rid of erroneous sequence variants which may distort the results. For each amino acid sequence $\sigma$ present in the data, we count how many one-mismatch variants are also present in the data, and denote that number $d(\sigma)$. For each $\sigma$ such that $d(\sigma)>2$, we generate all possible one-mismatch variants $\sigma'$ {\em in silico}, and calculate their $P_{\rm gen}(\sigma')$ using Monte Carlo simulations as described above. Finally, we calculate a p-value for each $\sigma$ corresponding to the probability that $\sigma$ has no less similar variants in null model than in the data, $\sum_{d'\geq d(\sigma)} P(d'|\sigma)$ using \eqref{eq:pois}. We correct p-values for multiple testing using Benjamini-Hochberg (BH) correction, and select clonotypes with BH-adjusted p-value $< 0.001$ as significant results (ALICE hits).

{\bf Including abundance information.} The basic pipeline only takes the occurence of nucleotide sequences in the sample into account, and not their abundance (read count). To include that information, we replace in the pipeline the number of similar sequences, $d$, by a sum of transformed abundances over these sequences, $s=\sum_{i=1}^d f(c_i)$, where $c_i$ is the abundance of the $i^{\rm th}$ nucleotide variant with similar amino acid sequence. There exists several choices for the transformation $f$. $f(c)=1-\delta_{c,0}$ gives back the basic method, $s=d$. $f(c)=c$ corresponds to summing the abundances of all similar variants, while $f(c)=\log(c)$ corresponds to summing their logarithms.
To define the null model, we assume that the abundance of each sequence is sampled at random from the distribution of empirical frequencies. Since this distribution followed a power law, we worked on the logarithmic scale and we picked $f(c)=\log(c)$.
To calculate $P(s|\sigma)$, under null hypothesis, we use the identity: $P(s|\sigma)=\sum_d P(s|d)P(d|\sigma)$, where $P(d|\sigma)$ was computed as described in the basic pipeline (Eq.~\ref{eq:pois}). Then, $P(s|d)$ is obtained as a $d$-fold convolution of $P_f(f)$, the probability distribution of the transformed abundances $f(c_i)$. For instance, $P(s|d=1)=P_f(f)$, $P(s|d=2)=\sum_f P_f(f)P_f(s-f)=(P_f*P_f)(s)$, etc., so that $P(s|d)={P_f^{(*)}}(s)$.
These quantities do not depend on $\sigma$ and are computed just once at the beginning of the procedure from the clonotype abundance distribution.
Applying this advanced version of ALICE to the yellow fever data of Fig.\,2a \cite{Pogorelyy2018_yf} yielded very similar results (Fig.\,S3) as the basic method. While it is slower to implement, the advanced method could still be useful because it is expected to be robust to wide ranges of repertoire sampling depths, while the basic version implicitly relies on many sequences not being captured by the sample.

{\bf Data and code availability. }
We used processed data provided by the authors of following papers: mixed lymphocyte reaction data from \cite{Emerson2014}; sorted na{\"\i}ve and effector memory repertoires from \cite{Thome2016}; repertoires of patients before and after anti-CTLA4 immunotherapy from \cite{Robert2014,Subudhi2016}; repertoires before and after immunization with live yellow fever vaccine from \cite{Pogorelyy2018_yf}; ankylosing spondylitis data from \cite{Komech2018}. For  \cite{Pogorelyy2018_yf} only the first replicate of each time point (day 0 and day 15) was used. For \cite{Subudhi2016} we used paired samples of patients before and after the first dose of therapy. 

Links to all datasets produced by Adaptive biotechnogies and also datasets from \cite{Pogorelyy2018_yf,Komech2018} used in this study are available at ALICE sample data github repository: \url{https://github.com/pogorely/ALICE_sample_data}.
All code for ALICE pipeline and particular analysis done for this paper is available at github repository: \url{https://github.com/pogorely/ALICE}.

{\bf Statistics. } To compare the normalized number of ALICE hits and maximum frequency of productive rearrangement between memory and na{\"\i}ve repertoires (Fig.\,1b) we used Wilcoxon rank-sum two-tailed test  ($N=52$ na{\"\i}ve subsets and $N=60$ memory repertoires, $p=2.3\times10^{-14}$ for maximum frequency of productive rearrangement, and $p=4.9\times10^{-16}$ for normalized number of ALICE hits). To compare the normalized number of ALICE hits in repertoires before and after MLR (Fig.\,1c) we used Wilcoxon rank-sum test ($N=12$ pre-MLR PBMC samples, $N=12$ MLR cultures, $p=7.4\times10^{-7}$). To compare number of ALICE hits with other statistics between pre- and post-treatment time points for immunotherapy patients we used Wilcoxon signed-rank two-tailed test: in Fig.\,2c ($N=21$ before, $N=21$ after therapy), the total number of clonotypes gave $p=0.0005$,  Shannon entropy gave $p=0.042$, and normalized ALICE hits gave $p=0.00024$; for Fig.\,S4 ($N=40$ before, $N=40$ after first dose of therapy), the total number of clonotypes gave $p=0.6$, Shannon entropy gave $p=0.6$, and normalized ALICE hits gave $p=0.0003$.

{\bf Estimating enrichment in certain amino acid position of the TCR motif.}
To estimate the enrichment of amino acids at specific positions in the set of expanded TCR,
we used a position weight matrix model of TCR selection \cite{Elhanati2014}. 
The sequence enrichment ratio takes a factorized form over amino acid positions, parametrized by selection coefficient $s_i(\sigma_i)$, where $\sigma_i$ denotes the amino acid of sequence $\sigma$ at position $i$. The predicted frequency in the expanded set is then:
\begin{equation}
P_{\rm sel}(\sigma)=\frac{1}{Z}P_{\rm gen}(\sigma)e^{\sum_{1}^{L}{s_i(\sigma_i)}} 
\end{equation}
where $Z$ a normalization factor.

The $s_i$ parameters were learned by gradient ascent of the likelihood function, to which an $L_2$ regularization term, $-\lambda\Vert s\Vert^2$, was added. Specifically selection coefficients are updated according to:
${s_{i}(\sigma_i)} \leftarrow {s_{i}(\sigma_i)}+\epsilon[P_{\rm data}(\sigma_i)-P_{\rm sel}(\sigma_i) -2\lambda{s_{i}(\sigma_i)}]$,
where $P_{\rm sel}(\sigma_i)$ is the predicted frequency of given amino acid at position $i$, and $P_{\rm data}(\sigma_i)$ is its observed frequency in the data.
After each  update, all $s_{i}(\sigma_i)$ are shifted by a common additive constant to satisfy following normalization constraint: $\sum_a P_{\rm gen}(\sigma_i)e^{s_i(\sigma_i)} = 1$.

We applied this inference procedure on the 26 sequences forming the central cluster of sequences from ankylosing spondylitis patients in Fig\,2d. $\epsilon$ was set to 0.5 and $\lambda$ was set to 0.02. The algorithm was initialized with $s_{i}(\sigma_i)=0$. The iterative procedure was repeated until the sum of the squared update difference was lower than $10^{-6}$.
The bottom logo of Fig.\,2e shows values of $s_{i}(\sigma_i)$ weighted by amino acid frequencies, so that the height of each letter is $P_{\rm data}(\sigma_i){s_{i}(\sigma_i)}$.

\section*{Acknowledgments}
This work was supported by the Russian Science Foundation grant n.15-15-00178.  M.V. Pogorelyy is supported by Skoltech Systems biology fellowship. This work was partially supported by the European Research Council Consolidator Grant n. 724208.  D.M. Chudakov and M. Shugay supported by grant of the Ministry of Education and Science of the Russian Federation Number 14.W03.31.0005 (in part of cancer immunotherapy data analysis). M. Shugay is supported by Russian Science Foundation grant n. 17-15-01495.

\bibliographystyle{pnas}

\begin{thebibliography}{10}

\bibitem{Heather2017}
Heather JM, Ismail M, Oakes T, Chain B
\newblock (2017) {High-throughput sequencing of the T-cell receptor repertoire:
  pitfalls and opportunities}.
\newblock \emph{Briefings in Bioinformatics} 19:bbw138.

\bibitem{Lindau2017}
Lindau P, Robins HS
\newblock (2017) {Advances and applications of immune receptor sequencing in
  systems immunology}.
\newblock \emph{Current Opinion in Systems Biology} 1:62--68.

\bibitem{Komech2018}
Komech EA, {et~al.}
\newblock (2018) {CD8+ T cells with characteristic T cell receptor beta motif
  are detected in blood and expanded in synovial fluid of ankylosing
  spondylitis patients.}
\newblock \emph{Rheumatology} 36:878--883.

\bibitem{Robert2014}
Robert L, {et~al.}
\newblock (2014) {CTLA4 blockade broadens the peripheral T-cell receptor
  repertoire}.
\newblock \emph{Clinical Cancer Research} 20:2424--2432.

\bibitem{Subudhi2016}
Subudhi SK, {et~al.}
\newblock (2016) {Clonal expansion of CD8 T cells in the systemic circulation
  precedes development of ipilimumab-induced toxicities}.
\newblock \emph{Proceedings of the National Academy of Sciences}
  113:11919--11924.

\bibitem{Pogorelyy2018_yf}
Pogorelyy MV, {et~al.}
\newblock (2018) {Precise tracking of vaccine-responding T-cell clones reveals
  convergent and personalized response in identical twins}.
\newblock \emph{bioRxiv/300343}.

\bibitem{DeWitt2015}
DeWitt WS, {et~al.}
\newblock (2015) {Dynamics of the cytotoxic T cell response to a model of acute
  viral infection}.
\newblock \emph{Journal of Virology} 249:JVI.03474--14.

\bibitem{Emerson2017}
Emerson RO, {et~al.}
\newblock (2017) {Immunosequencing identifies signatures of cytomegalovirus
  exposure history and HLA-mediated effects on the T cell repertoire}.
\newblock \emph{Nature Genetics} 49:659--665.

\bibitem{Faham}
Faham M, {et~al.}
\newblock (2017) {Discovery of T cell receptor $\beta$ motifs specific to
  HLA-B27-positive ankylosing spondylitis by deep repertoire sequence
  analysis}.
\newblock \emph{Arthritis {\&} Rheumatology} 69:774--784.

\bibitem{Pogorelyy2018_pgen}
Pogorelyy MV, {et~al.}
\newblock (2018) {Method for identification of condition-associated public
  antigen receptor sequences.}
\newblock \emph{eLife} 7:1--12.

\bibitem{Glanville2017}
Glanville J, {et~al.}
\newblock (2017) {Identifying specificity groups in the T cell receptor
  repertoire.}
\newblock \emph{Nature} 547:94--98.

\bibitem{Dash2017}
Dash P, {et~al.}
\newblock (2017) {Quantifiable predictive features define epitope-specific T
  cell receptor repertoires}.
\newblock \emph{Nature} 547:89--93.

\bibitem{Shugay2017}
Shugay M, {et~al.}
\newblock (2017) {VDJdb: a curated database of T-cell receptor sequences with
  known antigen specificity}.
\newblock \emph{Nucleic Acids Research} 46:419--427.

\bibitem{Venturi2006}
Venturi V, {et~al.}
\newblock (2006) {Sharing of T cell receptors in antigen-specific responses is
  driven by convergent recombination.}
\newblock \emph{Proc. Natl. Acad. Sci.} 103:18691--6.

\bibitem{Qi2016}
Qi Q, {et~al.}
\newblock (2016) {Diversification of the antigen-specific T cell receptor
  repertoire after varicella zoster vaccination}.
\newblock \emph{Science Translational Medicine} 8:332ra46--332ra46.

\bibitem{Venturi2011}
Venturi V, {et~al.}
\newblock (2011) {A mechanism for TCR sharing between T cell subsets and
  individuals revealed by pyrosequencing.}
\newblock \emph{J. Immunol.} 186:4285--4294.

\bibitem{Elhanati2018}
Elhanati Y, Sethna Z, Callan CG, Mora T, Walczak AM
\newblock (2018) {Predicting the spectrum of TCR repertoire sharing with a
  data-driven model of recombination.}
\newblock \emph{Immunological reviews} 284:167--179.

\bibitem{Madi2017}
Madi A, {et~al.}
\newblock (2017) {T cell receptor repertoires of mice and humans are clustered
  in similarity networks around conserved public CDR3 sequences}.
\newblock \emph{eLife} 6.

\bibitem{Murugan2012}
Murugan A, Mora T, Walczak AM, Callan CG
\newblock (2012) {Statistical inference of the generation probability of T-cell
  receptors from sequence repertoires.}
\newblock \emph{Proceedings of the National Academy of Sciences of the United
  States of America} 109:16161--6.

\bibitem{Marcou2017}
Marcou Q, Mora T, Walczak AM
\newblock (2018) {High-throughput immune repertoire analysis with IGoR}.
\newblock \emph{Nat. Commun.} 9:561.

\bibitem{Thome2016}
Thome JJC, {et~al.}
\newblock (2016) {Longterm maintenance of human naive T cells through in situ
  homeostasis in lymphoid tissue sites}.
\newblock \emph{Sci Immunol} 1:1--23.

\bibitem{Emerson2014}
Emerson RO, Mathew JM, Konieczna IM, Robins HS, Leventhal JR
\newblock (2014) {Defining the alloreactive T cell repertoire using
  high-throughput sequencing of mixed lymphocyte reaction culture}.
\newblock \emph{PLoS ONE} 9:1--7.

\bibitem{Dulphy1999}
Dulphy N, {et~al.}
\newblock (1999) {Common intra-articular T cell expansions in patients with
  reactive arthritis: identical beta-chain junctional sequences and
  cytotoxicity toward HLA-B27.}
\newblock \emph{Journal of immunology (Baltimore, Md. : 1950)} 162:3830--9.

\bibitem{May2002}
May E, {et~al.}
\newblock (2002) {Conserved TCR beta chain usage in reactive arthritis;
  evidence for selection by a putative HLA-B27-associated autoantigen.}
\newblock \emph{Tissue antigens} 60:299--308.

\bibitem{Elhanati2014}
Elhanati Y, {et~al.}
\newblock (2014) {Quantifying selection in immune receptor repertoires}.
\newblock \emph{Proceedings of the National Academy of Sciences of the United
  States of America} 111:9875--80.

\bibitem{Davis2011}
Davis MM, Altman JD, Newell EW
\newblock (2011) {Interrogating the repertoire: broadening the scope of
  peptide–MHC multimer analysis}.
\newblock \emph{Nature Reviews Immunology} 11:551--558.

\bibitem{DeWitt2018}
DeWitt WS, {et~al.}
\newblock (2018) Human t cell receptor occurrence patterns encode immune
  history, genetic background, and receptor specificity.
\newblock \emph{bioRxiv/313106}.

\bibitem{Sethna2018}
Sethna Z, Elhanati Y, Callan CG, Mora T, Walczak AM
\newblock (2018) {OLGA: fast computation of generation probabilities of B- and
  T-cell receptor amino acid sequences and motifs}.
\newblock \emph{bioRxiv/367904}.

\bibitem{Britanova2016}
Britanova OV, {et~al.}
\newblock (2016) {Dynamics of Individual T Cell Repertoires: From Cord Blood to
  Centenarians}.
\newblock \emph{The Journal of Immunology} 196:5005--5013.

\bibitem{Shugay2015}
Shugay M, {et~al.}
\newblock (2015) {VDJtools: Unifying Post-analysis of T Cell Receptor
  Repertoires}.
\newblock \emph{PLOS Computational Biology} 11:e1004503.

\end{thebibliography}

\setcounter{figure}{0}
\setcounter{table}{0}
\renewcommand{\thefigure}{S\arabic{figure}}
\renewcommand{\thetable}{S\arabic{table}}

\begin{figure}[p]
\noindent\includegraphics[width=\linewidth]{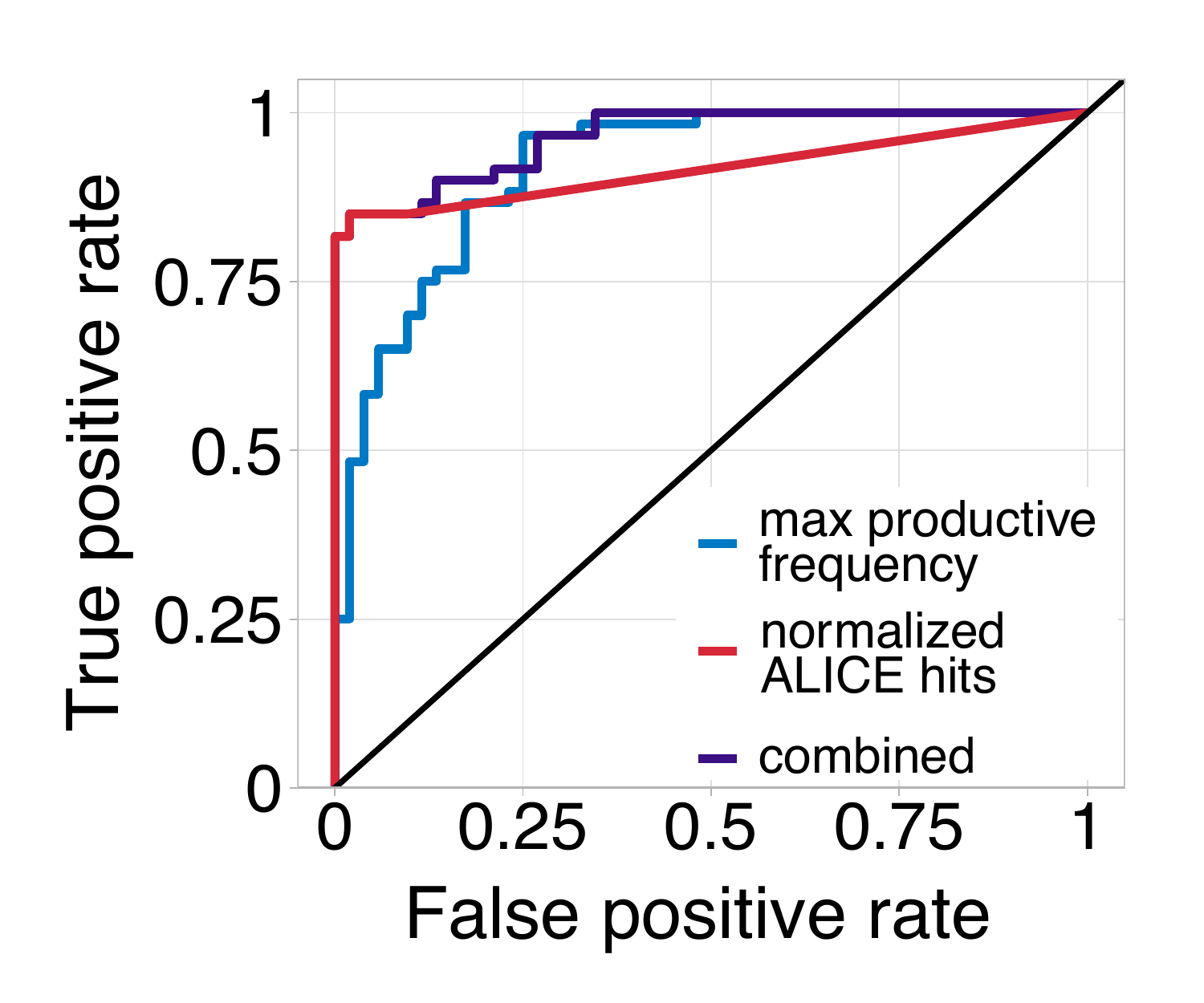}
\caption{{\bf Receiver operating characteristic (ROC) curves for classification of memory and na{\"\i}ve repertoires using ALICE hits and maximum productive rearrangement frequency as suggested in \cite{Thome2016}}. The classifier based on ALICE hits has a much higher true positive rate for low (up to 20\%) false positive levels, but it could not distinguish na{\"\i}ve and memory subpopulations both having zero hits. To break these ties, we ranked memory and na{\"\i}ve subsets with zero ALICE hits by the maximum frequency of productive rearrangements (combined classifier, purple curve). The areas under the ROC curve (AUROC) for these classifiers are 0.92 (ALICE hits-based), 0.92 (maximum productive frequency-based), 0.96 (combined classifier).}
\end{figure}

\begin{figure}[p]
\noindent\includegraphics[width=\linewidth]{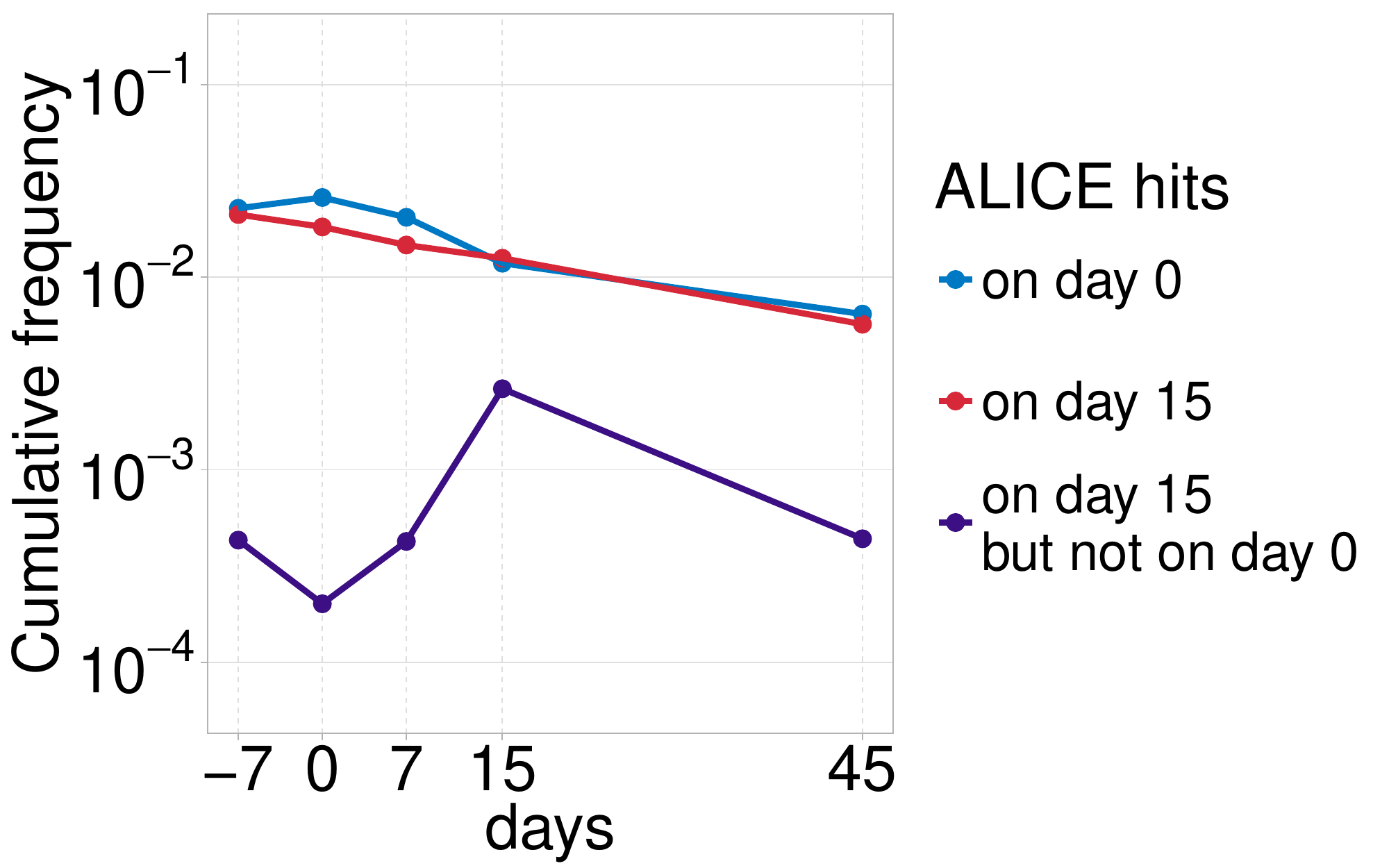}
\caption{{\bf Cumulative fraction of repertoire occupied by immune response signatures of donor Q1.} As we noted in the main text, one of the limitations of this approach is that it is hard to distinguish clonotypes specific for multiple conditions happening simultaneously, for instance response to vaccination and mild viral infection. Neither the signatures identified on day 0 nor the signatures identified on day 15, are able to recapitulate the dynamics of the yellow fever vaccine response. However, the subset of day 15 signatures that are absent on day 0 shows a clear yellow fever specific response with peak on day 15. The 122 clonotypes found as significant on both day 0 and day 15 are not similar (defined as one amino acid mismatch) to any the responding clonotypes as identified using temporal differences \cite{Pogorelyy2018_yf}, further suggesting that they are not yellow fever specific, but instead correspond to another immune response that is already contracting at day 0.
}
\end{figure}

\begin{figure}[p]
\noindent\includegraphics[width=\linewidth]{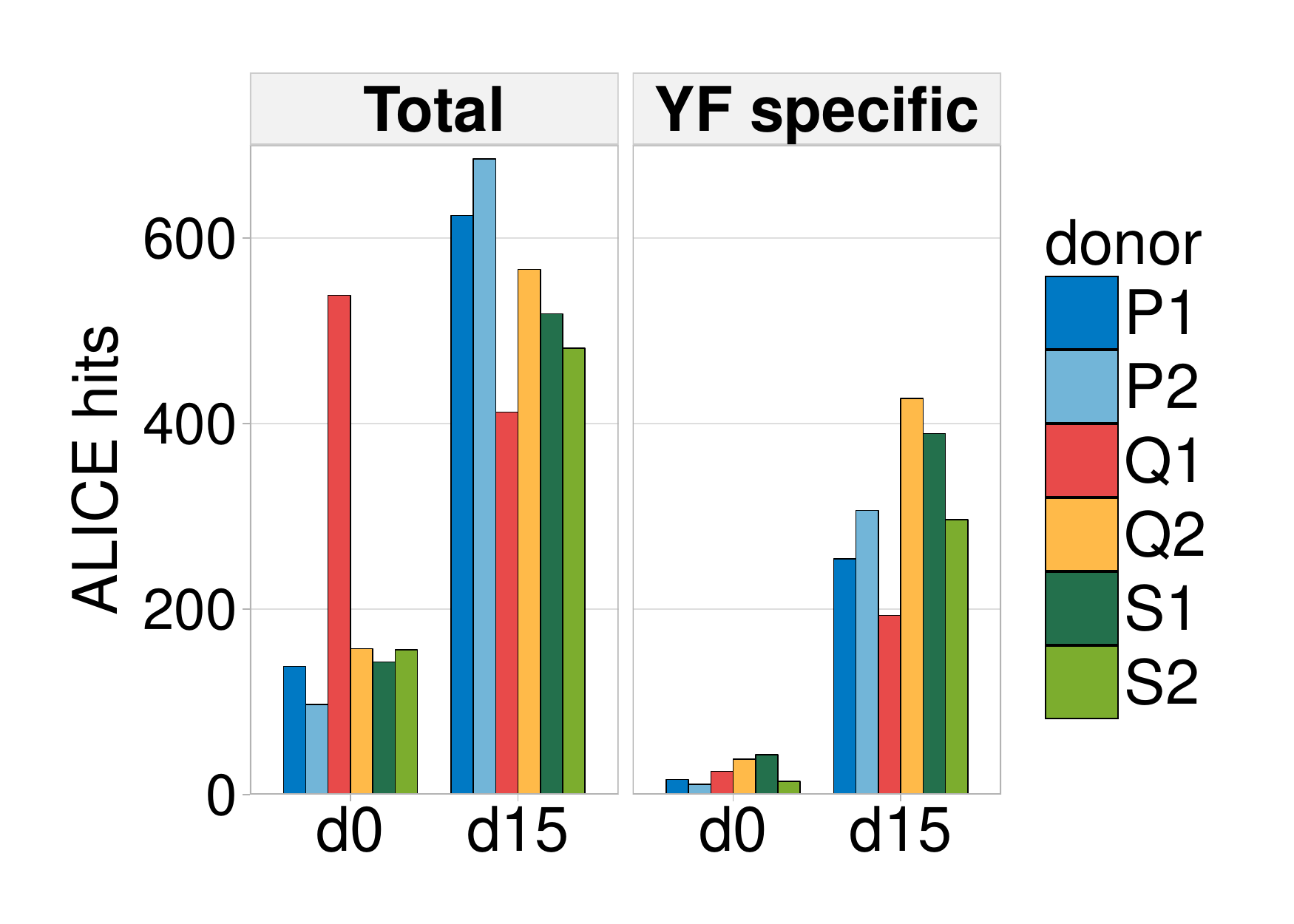}
\caption{{\bf Number of ALICE hits identified on day 0 and day 15 after yellow fever vaccination using abundance information.} The results of this analysis are almost identical to the results of Fig.~2a, with many more signatures identified after immunization than before (with the exception of donor Q1), and a large fraction of ALICE hits identified on day 15 having similar sequences to previously identified yellow fever specific clonotypes from \cite{Pogorelyy2018_yf}}.
\end{figure}

\begin{figure}[p]
\noindent\includegraphics[width=\linewidth]{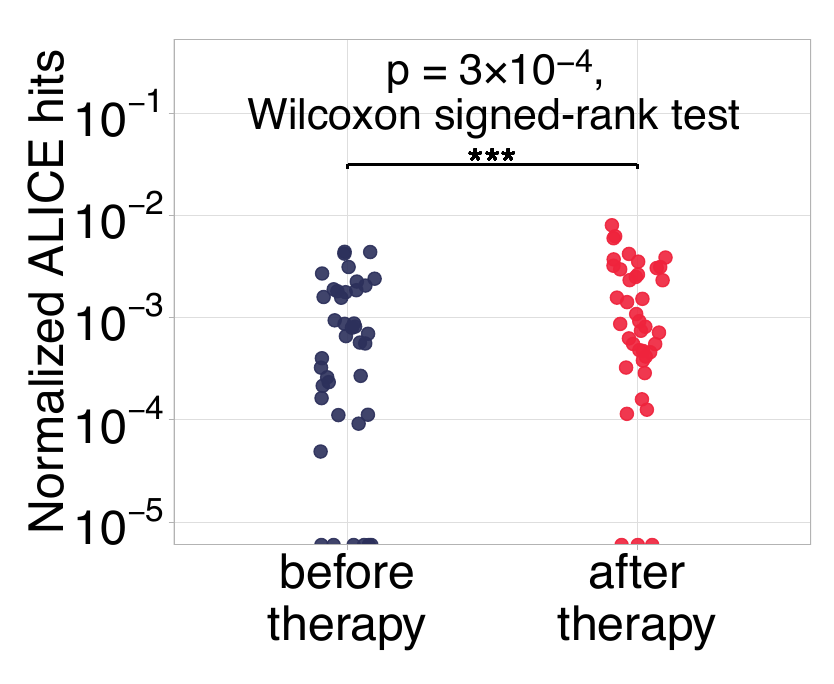}
\caption{{\bf Number of immune response signatures for \cite{Subudhi2016}.} The number of ALICE hits is significantly higher after immunotherapy than before.}
\end{figure}

\end{document}